\documentclass{article}

\usepackage{microtype}
\usepackage{graphicx}
\usepackage{subcaption}
\usepackage{booktabs} 

\usepackage{hyperref}

\usepackage[preprint]{icml2026}

\usepackage{algorithm}

\usepackage{algpseudocode}

\makeatletter
\algrenewcommand\alglinenumber[1]{\scriptsize #1:}
\makeatother

\newcommand{\StateCont}[1]{\Statex\hspace{\algorithmicindent}#1}

\usepackage{amsmath}
\usepackage{amssymb}
\usepackage{mathtools}
\usepackage{amsthm}

\usepackage[capitalize,noabbrev]{cleveref}

\icmltitlerunning{Security as a Learning Problem}

\begin{document}

\twocolumn[
  \icmltitle{Towards Adaptive, Learning-Based Security in Decentralized Applications}

 \begin{icmlauthorlist}
    \icmlauthor{Stefan Behfar}{cam}
    \icmlauthor{Jon Crowcroft}{cam}
  \end{icmlauthorlist}

  \icmlaffiliation{cam}{Department of Computer Science and Technology, University of Cambridge, Cambridge, United Kingdom}

  \icmlcorrespondingauthor{Stefan Behfar}{skb67@cam.ac.uk}

  \vskip 0.3in
]

  \icmlkeywords{Machine Learning, ICML}

  \vskip 0.3in
]



\printAffiliationsAndNotice{}  

\begin{abstract}
Web3 systems expose a fundamentally different security landscape from centralized platforms, characterized by composability, pseudonymous identities, decentralized governance, and rapidly evolving attack strategies that span social, application, and protocol layers. Existing security mechanisms—such as static smart contract analysis, blacklist-based phishing detection, and network-level mitigation—operate in isolation and assume fixed threat models, limiting their effectiveness against adaptive, cross-layer adversaries.
This position paper argues that securing Web3 requires a shift from static, tool-centric defenses to learning-driven security primitives capable of continuous reasoning, adaptation, and actuation. We introduce AI-powered smart certificates as a new security abstraction: programmable, continuously updated trust artifacts that integrate on-chain verifiability with off-chain machine learning signals derived from user behavior, transaction dynamics, and social context. Unlike traditional certificates or audits, these certificates maintain state, learn under distribution shift, and support automated policy enforcement and revocation in response to evolving threats.
We argue that existing paradigms—formal verification, threat modeling, and isolated anomaly detection—are structurally limited in capturing the non-stationary and socio-technical nature of Web3 attacks. We outline an architecture in which AI-powered smart certificates serve as cross-layer sentinels that coordinate heterogeneous security signals in real time, and position smart certificates as a research direction, raising questions around learning under partial observability, adversarial adaptation, and trustworthy ML deployment in decentralized systems.
\end{abstract}
 
\section{Introduction}
Blockchain Technology (BT) offers a framework of structural decentralization, paving the way for a fundamentally new approach to conducting business transactions, particularly among untrusted entities. The key components of BT—cryptography, transparency, and decentralized smart contracts and ledgers—support crucial functions such as verification, identification, authentication, integrity, and immutability. This technology facilitates the exchange of transactions within a large network of untrusted entities and maintains chronologically linked, duplicated digital ledgers in a decentralized database \cite{Maleh2020}.

Given that BT is a cutting-edge technology with a lot of potential, there are doubts about how reliable it is \cite{Krause2019}. If a system had such authority, then its most potent actors may collude to alter the blockchain or stop its contents from being broadcast. There have been numerous documented cyberattacks, and blockchain implementations have several cybersecurity flaws \cite{Luu2016}\cite{Taylor2019}. By transferring tiny quantities of "dust coins" to the personal wallets of the target victims, hackers were able to compromise the privacy and anonymity of Bitcoin (BTC) users, as evidenced by the attack on the Litecoin blockchain network in 2019. By undertaking the analysis of the addresses, hackers could discover the owner of each wallet and follow its global activity \cite{Omelchenko2019}. 

Web3, the third generation of the internet, is defined by decentralized applications and trustless interactions built on blockchain and other distributed ledger technologies. Unlike Web2 systems, Web3 lacks central intermediaries that can monitor, mediate, or respond to attacks—placing the burden of trust and verification on decentralized protocols and user interfaces \cite{behfar2023a,behfar2023b}.
Recent empirical studies \cite{Su2021,Chen2023} have revealed a broad range of attacks in Web3 ecosystems—from reentrancy and front-running on Ethereum-based contracts to large-scale manipulation on decentralized exchanges like Uniswap and Curve. These vulnerabilities are not isolated; they are increasingly polymorphic, behavioral, and context-aware, targeting the interplay of wallets, smart contracts, oracles, and APIs.

This position paper argues that current security paradigms—ranging from symbolic analyzers like MythX to curated blacklist-based tools like PhishTank—offer limited, siloed protection that cannot address the complexity and adaptability of real-world Web3 threats which are multi-dimensional (contract logic + social vectors + transaction dynamics) due to inability to adapt, inability to reason across user and contract layers, and lack of automated response or learning. AI, being structure-agnostic and adaptive, is uniquely capable of bridging this gap, extracting cross-layer patterns across user behavior, contract interactions, and transaction graphs. Furthermore, we advocate for a new security primitive: AI-Powered Smart Certificates—programmable digital agents capable of dynamic, semantic verification in decentralized ecosystems.
Our contributions are as follows:

\begin{itemize}

\item We argue that existing Web3 security paradigms — including formal verification, threat modeling, and isolated anomaly detection — are insufficient to address dynamic, cross-layer threats. We categorize key vulnerabilities such as phishing, smart contract exploits, and Sybil attacks, and demonstrate the structural failure of current approaches to mitigate these threats in a decentralized context.

\item We position AI as a necessary foundation for Web3 security due to its ability to perform adaptive, semantic, and behavioral analysis across application, wallet, and protocol layers. We systematize current uses of AI in anomaly detection and threat analysis and show where they fall short without integrated, actionable, real-time deployment.

\item We introduce and advocate for AI-driven Smart Certificates as a novel solution: dynamic, programmable certificates that integrate on-chain and off-chain intelligence. These certificates act as real-time sentinels capable of verifying context-aware behaviors, bridging siloed tools like MythX and OpenPhish, and enabling autonomous response in decentralized applications (dApps).

\end{itemize}

\section{Problem Setting and Threat Model}
Web3 enterprises encounter several challenges, including scalability issues due to the limitations of current blockchain technology, and security vulnerabilities such as smart contract exploits, phishing attacks, and social engineering. 
This has urged us to investigate a solution to help enterprises feel safe when using Web3 in their environment.

\subsection{Types of Attack to Web3 Applications}
In the realm of Web3, there are many attacks that can compromise the security of users' wallets:

\textbf{Social Media Originated Attacks:} \\
\textit{Phishing and Social Engineering Attacks: } These attacks exploit social platforms to deceive users into revealing sensitive information, such as private keys or login credentials. Attackers use fake profiles, messages, and links to trick users. For example, a scammer posing as a legitimate company representative on X (Twitter) directs users to a fake website to harvest their wallet credentials \cite{weinz_phishing2025, thomas_understanding}. Detection methods such as PhishTank \cite{phishtank} and OpenPhish provide datasets of known phishing URLs, which can be utilized to train AI-based phishing detection systems. These systems, employing NLP models such as BERT-based classifiers, are capable of detecting scam-related language in websites.

\textbf{Real-Time Activities:} \\
\textit{API Exploits: } Vulnerabilities in APIs used to integrate wallets with other platforms, including social media, can be exploited to access or manipulate wallet data. This can happen in real-time if APIs are not properly secured. For instance, an attacker intercepting API calls between a wallet and a social media platform to execute unauthorized transactions. Tools such as OWASP ZAP (Zed Attack Proxy) \cite{owasp_zap}, Postman, and Burp Suite could be utilized for detecting such activities.\\
\textit{Denial of Service (DoS) Attacks: } These attacks aim to overwhelm the blockchain network or wallet services with excessive requests, causing disruption. Though not always originating from social media, they can be coordinated through it. A coordinated attack from a social media platform that targets a wallet service with a flood of transaction requests, causing it to become unresponsive. Tools such as Cloudflare \cite{cloudflare_dos}, AWS Shield, and Snort \cite{snort_dos} could be used for detecting DoS attacks.

\textbf{Smart Contract Vulnerabilities}\\
Exploits targeting flaws in smart contracts can lead to significant financial losses. Tools like MythX \cite{mythx} and OpenZeppelin \cite{openzeppelin} are used to audit and secure smart contracts. For instance, a vulnerability in a DeFi contract allowing an attacker to drain funds by exploiting a flaw in the contract's code. Furthermore, tools  such as Slither \cite{slither} could also be used for detecting such vulnerabilities.

\textbf{Sybil Attacks}\\
Attackers create multiple fake identities within a decentralized network to gain undue influence or disrupt the network. This can impact voting, consensus mechanisms, and trust. For instance, an attacker creating numerous fake nodes to control a significant portion of a blockchain network, affecting its consensus process \cite{sybil_attacks_survey}. SybilGuard \cite{sybilguard}, SybilRank \cite{sybilrank} and Graph-Based Anomaly Detection \cite{graph_anomaly_detection} could be used for detecting Sybil attacks, which use graph-based neural networks or clustering on transaction graphs.

Other Attacks include: 51\% Attacks, Re-Entrancy Attacks, Front-Running Attacks, and Flash Loan Attacks.

\subsection{Why Existing Paradigms Fail: A Layered Threat Analysis}

Traditional security paradigms, rooted in centralized control, static threat boundaries, and predefined logic, struggle to address the dynamic and composable systems, see Table 1 for comparison of traditional vs. AI-based defenses. Furthermore, we discuss key defense strategies across four common security domains and expose their failure points in decentralized environments in Table 2.

\begin{table}[h]
\centering
\small
\caption{Comparison of Traditional vs. AI-Based Defenses}
\begin{tabular}{|p{1.5cm}|p{2.5cm}|p{3cm}|}
\hline
\textbf{Dimension} & \textbf{Traditional Tools} & \textbf{AI-Based Approach} \\
\hline
Scope & Layer-specific (e.g., contract-only) & Cross-layer (wallet + social + contracts) \\
\hline
Adaptability & Static rule-based signatures & Learns evolving patterns and behaviors \\
\hline
Data Modalities & Code or URLs & Contracts, behavior graphs, time series \\
\hline
Response & Manual, post-hoc & Real-time scoring + autonomous alerts \\
\hline
\end{tabular}
\end{table}

\begin{table}[ht]
\vspace{-1em}
\footnotesize
\centering
\caption{Systemic Mismatches Between Security Paradigms and Web3 Properties}
\label{tab:paradigm_mismatch}
\begin{tabular}{|p{1.5cm}|p{2.5cm}|p{3cm}|}
\hline
\textbf{Security Paradigm} & \textbf{Assumptions/Design Scope} & \textbf{Web3 Mismatch} \\
\hline
Formal Verification (e.g., MythX, Slither) & Assumes correctness under fixed logic and isolated contract scope. & Contracts are immutable post-deployment; runtime threats emerge. \\
\hline
Threat Modeling (e.g., STRIDE, DREAD) & Assumes well-defined, centrally governed system boundaries. & DAOs evolve via votes; attack surfaces shift based on protocol upgrades. \\
\hline
Anomaly Detection (e.g., PhishTank, OpenPhish) & Operates off-chain, detects statistical deviations in known patterns. & Lacks semantic understanding of contract logic; cannot generalize across diverse dApps and user behaviors. \\
\hline
Sybil Mitigation (e.g., SybilRank, SybilGuard) & Relies on stable social graphs and identity persistence. & Identities are cheap and ephemeral; trust signals are gameable in pseudonymous, incentive-driven settings. \\
\hline
\end{tabular}
\end{table}

Formal verification ensures a contract conforms to predefined specifications, typically via static analysis or symbolic execution. Tools such as MythX~\cite{mythx}, Slither~\cite{slither}, and OpenZeppelin~\cite{openzeppelin} serve as vital pre-deployment defenses. However, these tools cannot adapt to interactions with malicious external contracts, and do not protect against socio-technical threats like DAO proposal manipulation or phishing.
Conventional models such as STRIDE~\cite{STRIDE} and DREAD~\cite{dread_threat_modeling} classify threats via structured taxonomies. They perform well in systems with centralized control and known perimeters. In contrast, they struggle to generalize across novel dApps and multi-chain behaviors.
Behavioral detection systems based on transaction patterns have gained traction, such as PhishTank and OpenPhish. These typically focus on labeled phishing domains or transaction anomalies.
Also tools e.g. SybilRank~\cite{sybilrank} and SybilGuard~\cite{sybilguard} defend against identity-based attacks using trust graphs. These systems underperform in pseudonymous ecosystems where identities are disposable.

\subsection{Research Questions}

Despite advancements in existing tools and methodologies, this paper posits that the siloed nature of current Web3 security solutions fails to address the unique challenges of decentralized environments effectively. Current tools such as MythX, PhishTank, and Cloudflare operate independently, lacking integration into a cohesive framework capable of managing Web3’s multifaceted vulnerabilities. For instance, MythX excels in smart contract auditing, but it does not incorporate threat signals from external sources like phishing detection platforms (e.g., PhishTank). Similarly, Cloudflare is robust for DoS protection but lacks adaptability to Web3’s decentralized governance and diverse network structures, where rapid response mechanisms are critical.

\begin{figure}[ht]
    \centering
        \centering
        \includegraphics[width=0.48\textwidth]{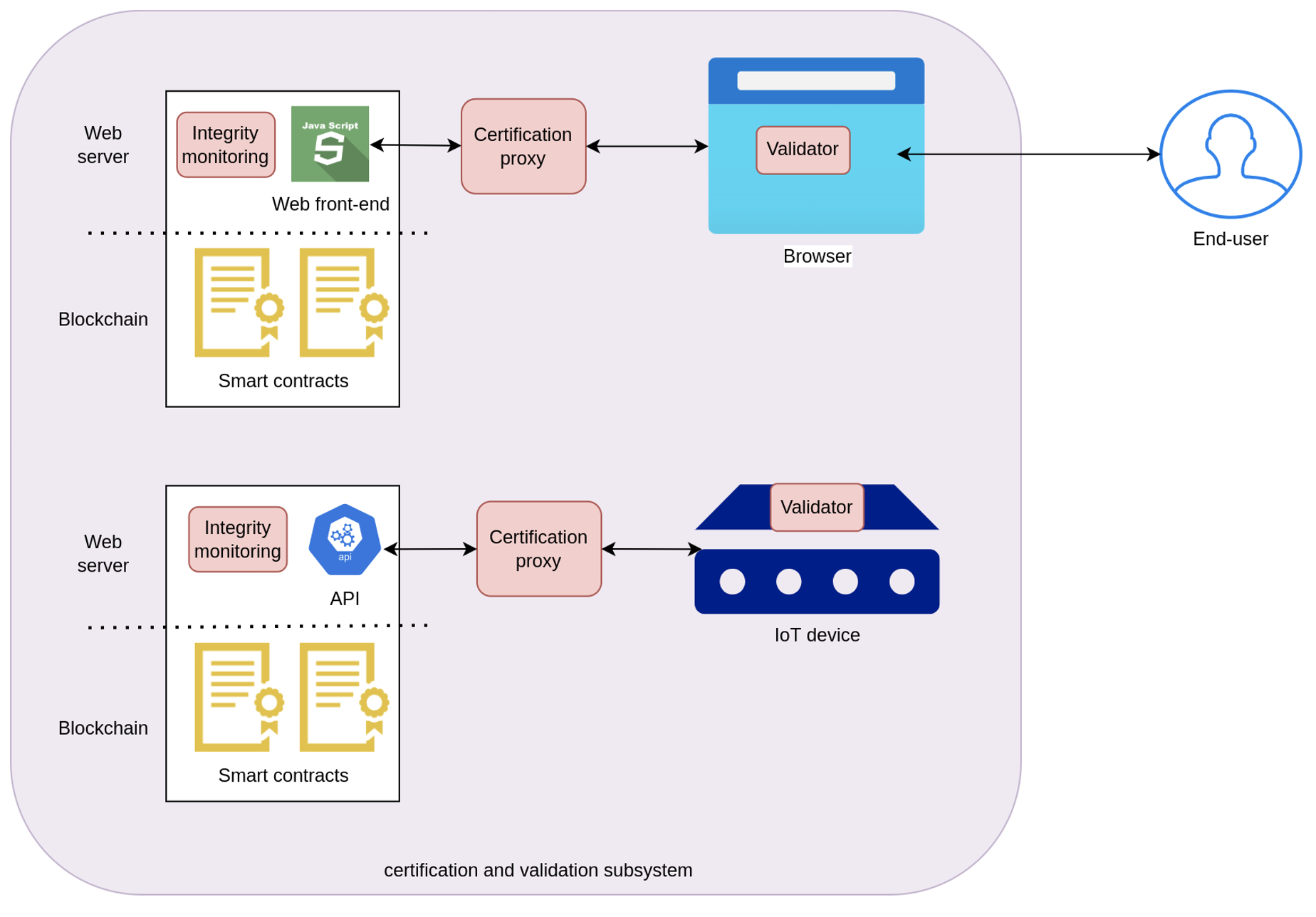}
        \caption{Certification and validation of Web applications.}
        \label{fig:certification}
\end{figure}
\begin{figure}[ht]
        \centering
        \includegraphics[width=0.48\textwidth]{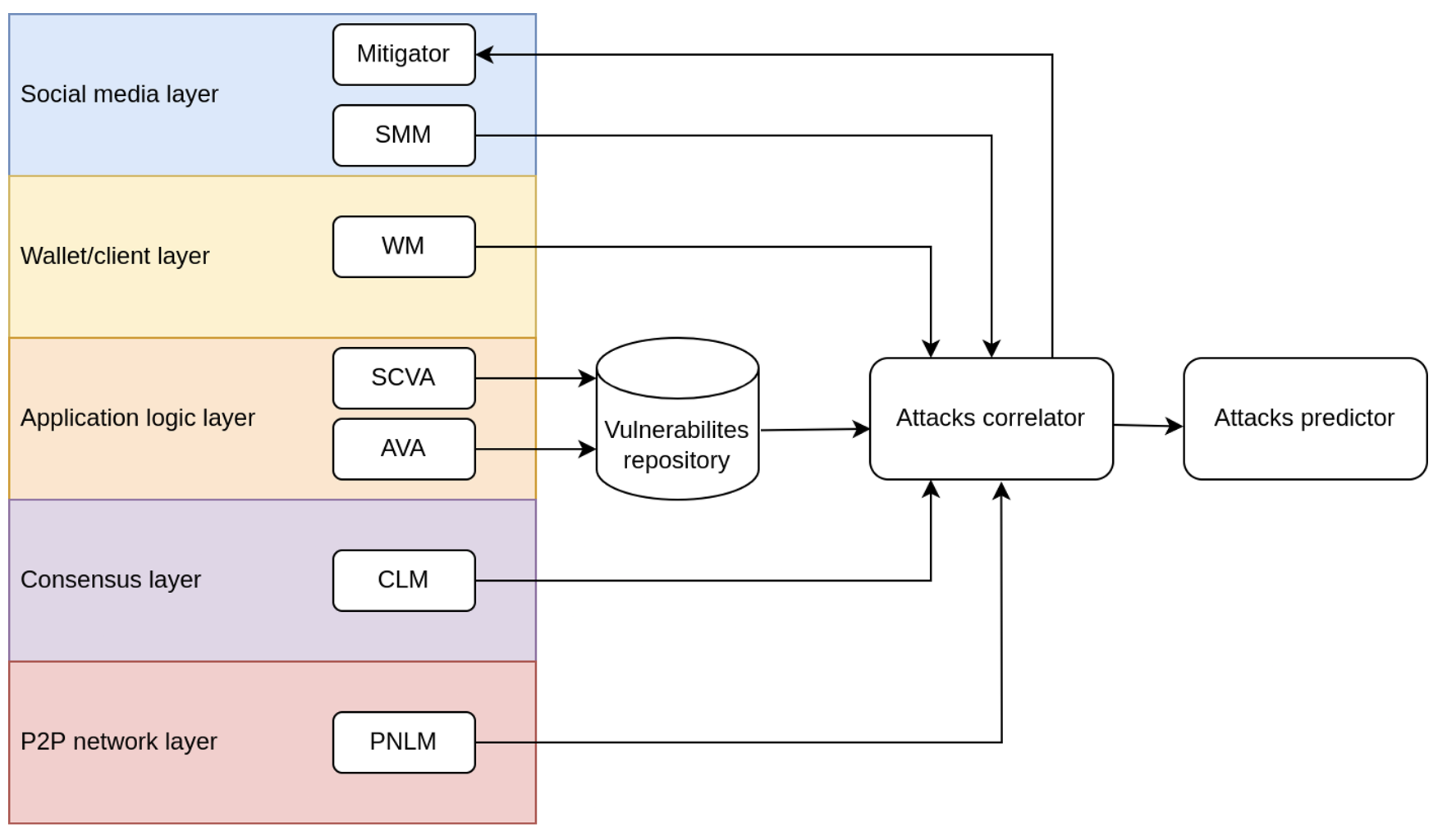}
        \caption{Web3-blockchain attacks to different layers, and attacks correlator.}
        \label{fig:attacks}
\end{figure}

To address the fragmented nature of current Web3 security tools, this paper proposes a user-centric cybersecurity reference architecture (Figure 1) that consolidates existing solutions into a unified framework specifically tailored for decentralized applications. Certification operates as intelligent middleware between end users (human or IoT) and decentralized applications, enabling adaptive, per-user security and privacy responses.
To accommodate varying user needs and application contexts, the architecture integrates user profiling. Lead users can define initial privacy and security preferences, which are then dynamically refined using AI-based analytics. This results in a continuously updated understanding of user behavior and risk exposure.

RQ1: How can AI components be designed to monitor and secure Web3 applications across diverse threat types?\\
The proposed smart certificates enhance the resilience of digital infrastructure by introducing cross-layer attack detection, identification, and mitigation mechanisms. This Web3–blockchain stack spans the media, wallet/client, and application layers—each integrated with tailored AI modules for detecting smart contract vulnerabilities, real-time cyberattacks, phishing attempts, and behavioral anomalies. AI techniques, such as graph-based anomaly detection and deep learning, offer the ability to capture polymorphic and multi-step attacks that rule-based systems fail to detect.

RQ2: How can smart certificates be generated and updated to mitigate vulnerabilities primarily caused by human factors?\\
Smart certificates must be adaptive and context-aware, supporting real-time updates informed by behavioral analytics, trust scoring, and usage patterns. Certificate generation incorporates insights from vulnerability detection tools (e.g., MythX, Slither), while updates reflect ongoing activity across the dApp ecosystem. These certificates can serve as intelligent agents that enforce best practices, block known risky interactions, and issue real-time alerts—mitigating social engineering, misconfigurations, and user error.

RQ3: How can response mechanisms be tailored to the user's security profile and threat context?\\
Effective threat response in Web3 requires contextual prioritization—ranking threats by likelihood and severity. The proposed architecture leverages user profiling to co-define response policies that align with specific needs (e.g., developer, investor, DAO participant). Security tools are dynamically reconfigured based on this profile to ensure usability and efficacy. Drawing from usable security principles, these personalized interventions can reduce friction while increasing security adherence.
\section{Solutions}   
Due to the increased complexity and variety of potential attack vectors in Web3 applications, more sophisticated and layered solutions are necessary to address the full scope of these security challenges. While there are existing tools and solutions that target specific layers and types of attacks, such as phishing, smart contract vulnerabilities, or denial-of-service attacks, none of these solutions offer a fully comprehensive approach across all layers of Web3 applications. We provide an architecture that integrates and leverages these existing tools into a more cohesive and robust solution.
This comprehensive architecture ensures security across various layers, including Wallet/Client, Media, and Application, each of which is vulnerable to distinct types of attacks. The typical structure of these layers in Web3 applications is illustrated in Figure 2, where the integration of real-time monitoring, threat detection, and mitigation across all layers is achieved by combining the strengths of existing tools and introducing AI-enhanced components for threat correlation and prediction.

\subsection{Integrated System Design}
To design and implement AI components that monitor and secure Web3 applications against various types of attacks, according to RQ1, one must integrate AI-driven threat detection and response mechanisms into the smart certificate framework. This involves developing algorithms that can analyze real-time data from social media, blockchain transactions, and API interactions to identify potential threats such as phishing, social engineering, smart contract exploits, API vulnerabilities, Sybil attacks, and DoS attacks. The AI components should leverage existing security tools like MythX for smart contract auditing and various API monitoring solutions to enhance threat detection capabilities. Upon detecting a threat, the AI system should generate a smart certificate that includes a detailed threat assessment and recommended mitigation steps, which can be installed on or redirected to the affected wallet address. This smart certificate acts as an automated security enforcer, providing users with real-time alerts and guidance to prevent and address vulnerabilities, thereby ensuring a comprehensive and proactive security posture for Web3 applications. 

For each layer (Figure 2) specific components will be provided to ensure robust evidence capturing, which thereafter is processed for identifying and predicting cybersecurity and risks. The resulting information then further goes through the cybersecurity decision tool-chain for automated mitigation and awareness strategies. see Figure 3 for an illustration of monitoring activity and attack detection, and attacks correlation subsystem inputted by the vulnerabilities’ repository, where the user spans the media and wallet layers.

With the traditional application architecture design, every application has to set up its own compatible servers for their own code execution. Additionally, such an approach is also implying a possibility for a single point of failure in the system. With the use of blockchain, each node (server, device etc.) on the network replicates the necessary data for all nodes, making a whole system more dynamic and reliable by putting up in front the true benefits of the decentralization approach. One could benefit from this concept by integrating the adaptation of the blockchain architecture in order to maximize the data security and integrity aspects of the proposed cybersecurity solution. 

\begin{figure*}
    \includegraphics[width=0.9\linewidth]{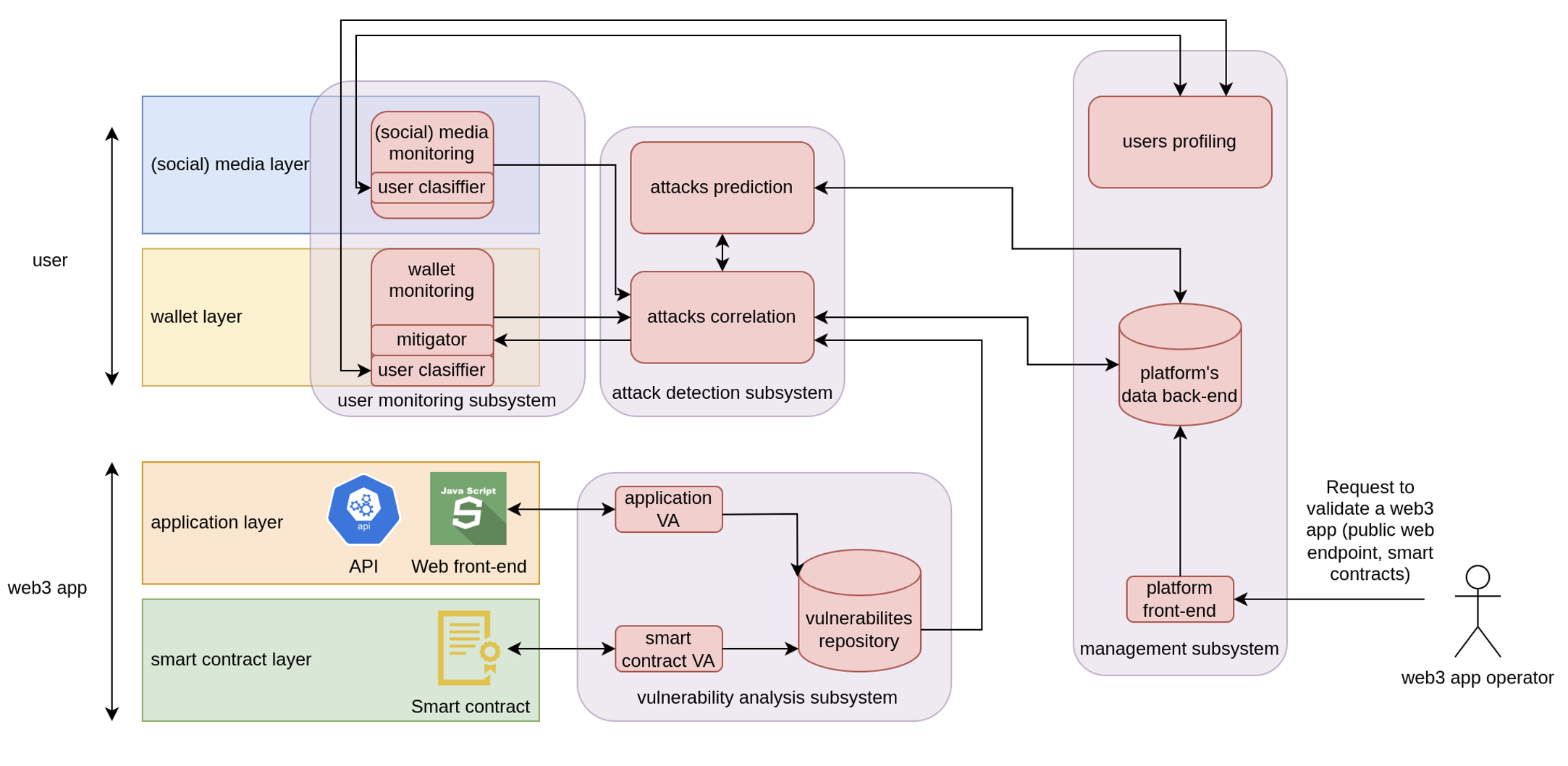} 
    \caption{Monitoring activity and attack detection, and attacks correlation subsystem inputted by the vulnerabilities’ repository, where the user spans the media and wallet layers. Attack correlation component integrates 1) social media monitoring via PhishTank/OpenPhish tool, 2) real-time activities’ monitoring for API exploits via OWASP ZAP and DoS attacks via Cloudflare tool, and 3) Application layer anomaly detection.} 
    \end{figure*}
    
In order to generate smart certificates, according to RQ2, for securing the Web3 applications against cyberattacks, the vulnerability scanning tool (shown in Figure 3) for securing the chain component will be integrated into the solution. The resulting framework will be efficient and resilient to various attacks, when the attack correlation component integrates 1) social media monitoring via PhishTank/OpenPhish tool, 2) real-time activities' monitoring for API exploits via OWASP ZAP and DoS attacks via Cloudflare tool, and 3) application layer anomaly detection and attack prediction via graph-based anomaly detection tools. 

\subsubsection{AI Methodologies Behind Smart Certificates}

The architecture operates by continuously monitoring each layer for signs of vulnerabilities. Data from each layer is aggregated in real-time and analyzed for potential threats. Once a vulnerability is detected, the system raises an alert and cross-references it with the vulnerability detection mechanisms.
To generate Smart Certificates for securing Web3 applications against cyberattacks (as posed in RQ2), our architecture integrates AI-driven components tailored to each system layer. The core framework combines time-series analysis, unsupervised clustering, and ensemble classification into a cross-layer security monitoring solution.

\textbf{AI-Enhanced Wallet Intelligence:}
Wallets are frequent targets for phishing and social engineering. To detect wallet-based anomalies:
\begin{itemize}
    \item PhishTank/OpenPhish APIs are employed to identify phishing URLs and malicious domains.
    \item A Long Short-Term Memory (LSTM) neural network monitors wallet behavior in real time, learning temporal patterns and detecting abrupt deviations, such as unauthorized token transfers or bot-driven interactions.
    \item Social media sentiment is analyzed using Transformer-based NLP models (e.g., DistilBERT) to detect high-risk language patterns and coordinated attack behavior.
    \item These detections trigger risk assessments and may lead to automatic credential revocation or real-time Smart Certificate updates.
\end{itemize}

\textbf{Real-Time API Monitoring:}
The media layer observes off-chain data streams and forums:
\begin{itemize}
    \item OWASP ZAP is used to inspect API endpoints with rule-based scanning, enhanced by AI-based anomaly detection.
    \item Unsupervised clustering algorithms (e.g., DBSCAN, Isolation Forest) analyze API usage patterns to detect anomalies and potential injection or privilege escalation attacks.
    \item Cloudflare's ML-powered DoS mitigation enhances defense against volumetric and protocol-level attacks.
\end{itemize}

\textbf{Application Layer Anomaly Detection:}
This layer ensures smart contract and API security:
\begin{itemize}
    \item MythX and OpenZeppelin audit smart contracts, feeding structured metadata into a classifier.
    \item A contract metadata classifier uses ensemble learning methods (e.g., Random Forest and XGBoost) to assign dynamic risk scores based on user ratings, audit findings, and operational behavior.
    \item Sybil attacks are addressed through graph-based methods such as SybilGuard and SybilRank, with potential for Graph Neural Network (GNN) integration in future iterations.
\end{itemize}

These combined AI techniques power the Smart Certificate engine, which remains unobtrusive and intervenes only when threats are deemed significant, aligning with the design goal described in RQ3.

\subsection{Smart Certificates}

Smart certificates act as real-time, programmable trust anchors within the proposed Web3 security architecture. They combine blockchain’s immutable auditability with advanced AI-driven threat detection to continuously assess and assure the security status of Web3 components—such as wallets, smart contracts, and client applications—throughout their operational lifecycle.

Certificates are issued following successful validation, such as passing security audits by established tools (e.g., MythX, Slither, or OpenZeppelin). These certificates signal that the component meets a defined set of security standards at the time of issuance. Importantly, this trust is revocable: if vulnerabilities are later detected—either through automated monitoring, updated threat intelligence, or adversarial simulation—the certificate can be invalidated. Wallet X cannot execute function Y on Contract Z because its AI-assigned score fell below threshold after a phishing pattern was detected.

Any entity interacting with a component (e.g., dApps querying wallets or contracts) can verify the certificate’s status on-chain. This mechanism ensures public verifiability, tamper resistance, and decentralized accountability, thereby aligning with the trustless ethos of Web3.
However, integrating AI into decentralized ecosystems introduces non-trivial challenges, such as transparency, resistance to model poisoning, and model interpretability. To mitigate these issues,
AI models may be hosted off-chain, but each prediction or decision can be accompanied by an on-chain cryptographic signature to preserve auditability. Differential privacy and federated learning techniques are considered to protect user data when learning from wallet or transaction behaviors.

\subsubsection{Implementation of Smart Certificates}

 Web3 Application provides a user interface that allows users to interact with the smart certificate system. Users can issue, verify, and manage certificates using this interface.
In this architecture, when an issuer issues a smart certificate through Smart Contract 1, the certificate metadata (e.g., recipient's name, issuer's name, date of issuance) and the cryptographic hash of the certificate document are recorded on the blockchain. Recipients or third parties can then verify the certificate's authenticity by checking the hash on Smart Contract 2. If a certificate needs to be revoked, Smart Contract 3 handles the revocation process. 
The formulas for smart certificates in Web3 involve various cryptographic operations and algorithms. Here, we'll outline the main formulas used in the issuance, verification, and revocation of smart certificates which include the AI components:

\textbf{Smart Contract (Certificate Issuance):} This smart contract handles the process of issuing smart certificates. Issuers interact with this contract to provide the necessary information and credentials for certificate issuance.

1. Define the smart certificate data structure:
    \begin{equation}
        \begin{split}
            \text{Certificate}: \{ & \text{ID, Recipient's Name,} \\
            & \text{Issuer's Name, ...} \}
        \end{split}
    \end{equation}
2. Generate a digital signature using the issuer's private key (\texttt{SKi}) for the certificate document (\texttt{d}) hash:
    \begin{equation}
    \text{Signature}: \text{Sigi} = \text{Sign}(\text{SKi}, \text{Hash}(d)) 
    \end{equation}
3. Store the certificate metadata and digital signature on the blockchain:
    \begin{equation}
        \begin{aligned}
        \text{Certificate On-Chain Data}: \{ \text{ID, Recipient's} \\
        \text{Name, Issuer's Name, Metadata, Signature} \}
        \end{aligned}
        \end{equation}

\textbf{Smart Contract (Certificate Verification):} This smart contract is responsible for verifying the authenticity and validity of a smart certificate. Recipients or third parties can interact with this contract to check the cryptographic hash of the certificate stored on the blockchain for verification.

1. Retrieve the certificate from the blockchain with its on-chain data.

2. Verify the digital signature using the issuer's public key (\texttt{PKi}) and the certificate document (\texttt{d}) hash:
    \begin{equation}
    \text{Verification}: \text{Verify}(\text{PKi}, \text{Hash}(d), \text{Sigi}) 
    \end{equation}

\textbf{Smart Contract (Certificate Revocation):}
A core component of the smart certificate lifecycle is the
ability to revoke certificates when security or policy
violations are detected.
Revocation may be triggered by a variety of conditions,
including fraudulent behavior, detected anomalies,
policy non-compliance, or certificate expiration.
In the proposed design, revocation is integrated directly into
the certificate state management logic.

Let's create a step-by-step example of implementing a smart certificate.

We consider a Solidity smart contract that defines a
Certificate of Deposit (CD) for use in a decentralized
exchange (DEX) environment.
The contract governs the issuance, verification, and
revocation of certificates, and can be integrated into a DEX
to allow users to issue, verify, and potentially trade CD
tokens representing ownership of certified assets or
behaviors.
While such certificates provide a foundational trust
primitive, their static nature limits their effectiveness
against evolving threats in Web3 systems.

\noindent
To enhance security, smart certificates should be augmented
with vulnerability analysis, anomaly detection, or attack
prediction capabilities driven by off-chain data and machine
learning models (Figure~3).
However, integrating learning-based security mechanisms with
smart contracts introduces fundamental challenges, as
contracts cannot directly access external data sources or
invoke machine learning services.
These constraints motivate a design in which learning-based
inference is performed off-chain, while verification and
enforcement remain on-chain.

\noindent
Before detailing the update process, we view a smart
certificate as a stateful on-chain object whose validity is
continuously assessed using off-chain signals.
The certificate state consists of risk and anomaly scores,
revocation status, and cryptographic commitments to the model
and evidence used for inference.
This state evolves according to a learning-based update rule
driven by off-chain analysis and enforced on-chain, with
revocation and enforcement acting as protocol-level
actuation points.

\noindent\textbf{Step I: Oracle/relayer-based anomaly inference.}
While the smart contract defines the lifecycle of certificates on-chain, it cannot
directly invoke external machine learning services or issue outbound HTTP requests.
To enable learning-based security analysis, the system therefore relies on an
oracle/relayer pattern, formalized in Algorithm~\ref{alg:smartcert-oracle}.

\noindent
When a certificate $c$ requires evaluation, the contract emits an
\texttt{InferenceRequested} event containing a unique request identifier $r$,
the certificate identifier, a commitment $fh=\mathsf{H}(x)$ to the off-chain
feature vector $x$, and a Merkle root $er$ committing to a batch of supporting
evidence.
The feature vector may include URL intelligence, transaction-graph statistics,
wallet behavior metrics, or other off-chain security signals relevant to the
application context (e.g., a decentralized exchange).

\noindent
An off-chain relayer/oracle observes the emitted event, reconstructs the feature
vector $x$, verifies the commitment $fh$, and applies the machine learning model
$\mathsf{ML}$ to compute a risk score $\mathit{score}$, a binary anomaly indicator
$\mathit{isAnom}$, and an enumerated threat classification $\mathit{reason}$.
For auditability, the oracle also produces a \emph{model commitment}
$\mathit{mc}$ (e.g., a cryptographic hash of the model weights or version).
If the inferred risk exceeds predefined policy thresholds, or if the threat
classification belongs to a critical category, the oracle assigns a nonzero
revocation timestamp $\mathit{revokedAt}$.

\noindent
The oracle submits the signed inference result back to the smart contract via
\texttt{SubmitInference}, together with a Merkle inclusion proof $(leaf,\pi)$
linking the certificate to the committed evidence root $er$.
The contract verifies the oracle signature and the Merkle proof before accepting
the update, ensuring that inference remains off-chain while preserving on-chain
verifiability and accountability.

\noindent\textbf{Step II: Verifiable certificate state update.}
Upon successful verification of the oracle response, the smart contract updates
the certificate state on-chain as specified in Algorithm~\ref{alg:smartcert-oracle}.
The inferred risk score, anomaly indicator, and threat classification are recorded
by updating the fields \texttt{riskScore}, \texttt{isAnomaly}, and
\texttt{reason}, respectively.

\noindent
To ensure accountability and reproducibility, the contract stores the
\texttt{modelCommitment} associated with the inference, binding the security
decision to a specific machine learning model or version.
The Merkle root \texttt{evidenceRoot} anchors the off-chain evidence batch used
during inference, enabling third parties to verify that a particular piece of
evidence was included without revealing the raw data on-chain.

\noindent
If a revocation condition is met, the \texttt{revokedAt} timestamp is set and cannot
be reverted, enforcing monotonic security decisions.
The \texttt{lastUpdateAt} field records the block timestamp of the most recent
evaluation.
All inference requests and updates are logged through the
\texttt{InferenceRequested} and \texttt{InferenceCommitted} events, providing an
auditable trail of security-relevant state transitions.

.

\begin{algorithm}[H]
\caption{Oracle/Relayer-based smart certificate updates with model commitment and Merkle evidence linkage (Part I: On-chain)}
\label{alg:smartcert-oracle}
\begin{algorithmic}[1]
\Require Smart certificate contract $\mathsf{CertSC}$; trusted oracle public key $pk_R$
\Require Hash function $\mathsf{H}(\cdot)$; signature verification $\mathsf{Verify}(\cdot)$; Merkle verification $\mathsf{MerkleVerify}(\cdot)$

\Statex \textbf{Certificate state (per certificate $c$):}
\State $\mathsf{cert}[c].riskScore \in [0,10000]$\\ $\mathsf{cert}[c].isAnomaly \in \{0,1\}$;\; $\mathsf{cert}[c].reason \in \mathbb{N}$
\State $\mathsf{cert}[c].revokedAt \in \mathbb{N}$ (0 if active)\\ $\mathsf{cert}[c].modelCommitment \in \{0,1\}^{256}$\\ $\mathsf{cert}[c].evidenceRoot \in \{0,1\}^{256}$
\State $\mathsf{cert}[c].lastUpdateAt \in \mathbb{N}$

\Statex \textbf{Pending request metadata (per request $r$):}
\State $\mathsf{pending}[r] \in \{0,1\}$;\;\; $\mathsf{reqMeta}[r] = (c, fh, er)$
\Statex \Comment{$fh$: hash of off-chain feature vector; $er$: Merkle root committing to evidence batch}

\Procedure{RequestInference}{$c, fh, er$}
  \State $r \gets \mathsf{H}(c, fh, er, block.number, \mathsf{nonce})$
  \State $\mathsf{pending}[r] \gets 1$
  \State $\mathsf{reqMeta}[r] \gets (c, fh, er)$
  \State \textbf{emit} $\mathsf{InferenceRequested}(r, c, fh, er)$
  \State \Return $r$
\EndProcedure

\Procedure{SubmitInference}{$r$}
  \Statex \textbf{Input:} $score,\, isAnom,\, revokedAt,\, reason,\, mc,$
  \StateCont{$leaf,\, \pi,\, sig$}
  \State \textbf{require} $\mathsf{pending}[r]=1$
  \State $(c, fh, er) \gets \mathsf{reqMeta}[r]$

  \State $h \gets \mathsf{H}(r, score, isAnom, revokedAt,$
  \State{$\qquad\quad reason, mc, er)$}
  \State \textbf{require} $\mathsf{Verify}(pk_R, h, sig)$
  \State \textbf{require} $\mathsf{MerkleVerify}(leaf, \pi, er)=\texttt{true}$

  \State $\mathsf{pending}[r] \gets 0$
  \State $\mathsf{cert}[c].riskScore \gets score$
  \State $\mathsf{cert}[c].isAnomaly \gets isAnom$
  \State $\mathsf{cert}[c].reason \gets reason$
  \State $\mathsf{cert}[c].modelCommitment \gets mc$
  \State $\mathsf{cert}[c].evidenceRoot \gets er$
  \State $\mathsf{cert}[c].lastUpdateAt \gets block.timestamp$

  \If{$revokedAt > 0$}
    \State $\mathsf{cert}[c].revokedAt \gets revokedAt$ \Comment{irreversible once set}
  \EndIf

  \State \textbf{emit} $\mathsf{InferenceCommitted}($
  \State{$\qquad\quad r,\, c,\, score,\, isAnom,$}
  \StateCont{$\qquad\quad \mathsf{cert}[c].revokedAt,\, reason,\, mc,\, er)$}
\EndProcedure
\end{algorithmic}
\end{algorithm}

\begin{algorithm}[t]
\ContinuedFloat
\caption{Oracle/Relayer-based smart certificate updates (continued) (Part II: Off-chain relayer/oracle)}
\begin{algorithmic}[1]
\Require Off-chain relayer/oracle $\mathsf{R}$ with signing key $sk_R$
\Require ML service $\mathsf{ML}$; evidence store $\mathsf{E}$; thresholds $\tau_{\mathrm{revoke}}$, critical reasons $\mathcal{R}_{\mathrm{critical}}$

\Statex \textbf{Relayer loop (off-chain):}
\While{listening to $\mathsf{InferenceRequested}(r, c, fh, er)$ events}
  \State $x \gets \mathsf{FetchFeatures}(c)$ \Comment{URLs, tx-graph stats, wallet actions, social signals, \dots}
  \State \textbf{assert} $\mathsf{H}(x)=fh$
  \State $mc \gets \mathsf{GetModelCommitment}(\mathsf{ML})$ \Comment{e.g., hash(model weights/version)}

  \State $(score, isAnom, reason) \gets \mathsf{ML}(x)$
  \If{$score < \tau_{\mathrm{revoke}}$ \textbf{or} $reason \in \mathcal{R}_{\mathrm{critical}}$}
    \State $revokedAt \gets \texttt{now}$
  \Else
    \State $revokedAt \gets 0$
  \EndIf

  \State $(leaf,\pi) \gets \mathsf{E.GetInclusionProof}(er, c)$
  \State $h \gets \mathsf{H}(r, score, isAnom, revokedAt, reason, mc, er)$
  \State $sig \gets \mathsf{Sign}(sk_R, h)$

  \State \textbf{call} $\mathsf{CertSC.SubmitInference}($
  \State{$\qquad\quad r,\, score,\, isAnom,\, revokedAt,$}
  \StateCont{$\qquad\quad  reason,\, mc,\, leaf,\, \pi,\, sig)$}
\EndWhile
\end{algorithmic}
\end{algorithm}

\section{Conclusion}

The decentralized and trustless nature of Web3 applications introduces complex, multi-layered security challenges that differ fundamentally from those of traditional, centralized systems. High-profile incidents such as the DAO exploit and the Poly Network hack have illustrated how vulnerabilities at different layers—ranging from smart contracts to API endpoints—can be exploited in concert, revealing the insufficiency of siloed, rule-based security tools.
This position paper contributes a structured understanding of Web3 security risks and highlights the fragmented nature of current detection and mitigation tools, which typically address isolated threats but fall short of offering cross-layer, adaptive defense mechanisms. By profiling vulnerabilities across the Wallet/Client, Application, and Smart Contract layers, we emphasize the need for a holistic security architecture.
Our proposed AI framework integrates predictive modeling, user behavior analysis, and real-time contract scoring to protect Web3 ecosystems. Unlike static tools, such a learning-based approach can evolve alongside attacker strategies and support context-aware defense capabilities.\\
\textbf{Research Outlook.}
Rather than presenting a finalized solution, this paper aims to motivate a broader research direction, outlining a \emph{Research Agenda: Learning-Based Security in Decentralized Systems} in the Appendix.

\section*{Impact Statement}

This position paper advocates for learning-based security mechanisms in decentralized systems, with a particular focus on AI-powered smart certificates for Web3 applications. The primary goal of this work is to advance machine learning research on adaptive, trustworthy decision-making under non-stationary, adversarial, and partially observable environments.

The potential positive societal impact of this work includes improved security and resilience of decentralized digital infrastructure, which may reduce large-scale financial losses, mitigate fraud and phishing attacks, and increase user trust in open, permissionless systems. By framing security as a learning problem, this work may also encourage the development of more robust, transparent, and continuously improving ML-based defenses applicable beyond Web3, including in IoT ecosystems and decentralized digital services.

At the same time, the use of AI-driven trust scoring and automated policy enforcement raises important ethical considerations. Incorrect or biased learning signals could lead to false positives, unjustified access restrictions, or exclusion of benign participants. Additionally, automated certificate revocation and enforcement mechanisms may concentrate decision-making power in ML models, potentially conflicting with principles of decentralization and user autonomy if not carefully governed. Adversarial manipulation, model opacity, and data privacy risks are also relevant concerns in decentralized settings.

This paper does not propose immediate large-scale deployment, but rather aims to stimulate discussion and future research on how such systems can be designed with appropriate safeguards, transparency, human oversight, and governance mechanisms. We believe that explicitly acknowledging these risks is essential for responsible progress in learning-based security systems, and we encourage future work to address robustness, interpretability, and accountability as first-class objectives.


\bibliography{references}
\bibliographystyle{icml2026}

\newpage
\appendix
\onecolumn
\section{Research Agenda: Learning-Based Security in Decentralized Systems}

This position paper advocates for AI-powered smart certificates as a new security primitive for Web3. Beyond the concrete architecture outlined earlier, this perspective raises a broader set of open \emph{machine learning research questions} that extend beyond blockchain security. We argue that decentralized security should be treated as a \emph{learning problem under uncertainty}, and we outline key research directions at the intersection of machine learning, security, and decentralized systems.

\subsection{Security as Online Learning Under Distribution Shift}

Web3 adversaries continuously adapt their strategies by exploiting new contracts, social narratives, and economic incentives. As a result, security signals are inherently non-stationary, rendering static threat models and one-time audits insufficient. This setting naturally motivates security mechanisms that operate as \emph{online learners}, continuously updating trust assessments and policies as new data arrives.

Open research questions include:
\begin{itemize}
    \item How can smart certificates learn robust trust scores under persistent distribution shift and delayed feedback?
    \item What notions of regret, stability, or calibration are meaningful when adversaries actively manipulate data distributions?
    \item How should memory, forgetting, and update schedules be designed in adversarial and decentralized environments?
\end{itemize}

\subsection{Learning Under Partial Observability and Cross-Layer Uncertainty}

In decentralized systems, no single entity has access to complete system state. Observations are fragmented across wallets, smart contracts, social platforms, and off-chain services. Smart certificates must therefore reason under \emph{partial observability}, integrating heterogeneous and potentially unreliable signals.

This raises several ML challenges:
\begin{itemize}
    \item Multimodal representation learning with missing, delayed, or noisy inputs
    \item Inferring latent threat states from sparse, cross-layer observations
    \item Robust inference when signals may be adversarially spoofed or selectively withheld
\end{itemize}

These problems connect naturally to research in probabilistic modeling, representation learning, and partially observable decision processes.

\subsection{Adversarial Adaptation and Game-Theoretic Learning}

Web3 security involves strategic adversaries who adapt in response to deployed defenses. Smart certificates introduce a feedback loop in which attacker behavior influences certificate scores, which in turn affect access control and execution policies.

Key research directions include:
\begin{itemize}
    \item Modeling attacker--defender interactions as repeated or stochastic games
    \item Studying equilibrium behavior when defenses themselves are learning-enabled
    \item Designing learning algorithms that remain robust under adversarial manipulation and data poisoning
\end{itemize}

This situates smart certificates within broader work on adversarial machine learning, online optimization, and game-theoretic learning.

\subsection{Governance-Aware and Human-in-the-Loop Learning}

Security decisions in Web3 are not purely technical; they are mediated by decentralized governance, social consensus, and human oversight. Certificate revocation, threshold selection, and policy enforcement may be contested or require justification.

Open challenges include:
\begin{itemize}
    \item Incorporating human feedback and governance signals into learning objectives
    \item Designing interpretable trust scores and explanations suitable for decentralized decision-making
    \item Balancing automated enforcement with human override in high-stakes scenarios
\end{itemize}

These challenges intersect with research on interpretability, human-in-the-loop learning, and value alignment.

\subsection{Trustworthy and Auditable Machine Learning in Decentralized Environments}

Deploying ML models in trustless settings introduces new requirements around auditability, verifiability, and accountability. Smart certificates highlight the need for ML systems whose outputs can be publicly inspected and contested.

Promising research directions include:
\begin{itemize}
    \item Verifiable ML inference and cryptographic attestations of model outputs
    \item Privacy-preserving learning from sensitive behavioral data
    \item Federated and decentralized learning under adversarial participation
\end{itemize}


\end{document}